\documentclass[letterpaper]{article} 
\usepackage{aaai23}  
\usepackage{times}  
\usepackage{helvet}  
\usepackage{courier}  
\usepackage[hyphens]{url}  
\usepackage{graphicx} 
\usepackage{natbib}  
\usepackage{caption} 
\usepackage{algorithmic}
\usepackage{subfig}
\usepackage{float}
\usepackage[switch]{lineno}
\usepackage{amsmath,amsfonts}

\usepackage{booktabs}
\usepackage{amsmath,lipsum}
\usepackage{cuted}
\usepackage{color} 
\usepackage{multirow}
\usepackage{amsmath, bm}
\usepackage{graphicx}
\usepackage{epstopdf}
\usepackage{bbding}
\usepackage{cite}
\usepackage[linesnumbered,ruled,vlined]{algorithm2e}
\usepackage{balance}
\usepackage{amsmath}
\usepackage{listings}
\usepackage{enumerate}
\usepackage{enumitem}
\nocopyright

\title{Tracing the Origin of Adversarial Attack for Forensic Investigation and Deterrence}
\author {
    Han Fang\textsuperscript{\rm 1},
    Jiyi Zhang \textsuperscript{\rm 1},
    Yupeng Qiu \textsuperscript{\rm 1},
    Ke Xu \textsuperscript{\rm 2},
    Chengfang Fang \textsuperscript{\rm 2},
    Ee-Chien Chang \textsuperscript{\rm 1}\footnote{Corresponding Authors.}
}
\affiliations {
    \textsuperscript{\rm 1} National University of Singapore\\
    \textsuperscript{\rm 2} Huawei International\\
    fanghan@nus.edu.sg, jiyizhang@u.nus.edu, qiu$\_$yupeng@u.nus.edu, changec@comp.nus.edu.sg 
}

\begin{document}
\maketitle

\begin{abstract}
Deep neural networks are vulnerable to adversarial attacks. In this paper, we take the role of investigators who want to trace the attack and identify the source, that is, the particular model which the adversarial examples are generated from. Techniques derived would aid forensic investigation of attack incidents and serve as deterrence to potential attacks.
We consider the buyers-seller setting where a machine learning model is to be distributed to various buyers and each buyer receives a slightly different copy with same functionality. A malicious buyer generates adversarial examples from a particular copy $\mathcal{M}_i$ and uses them to attack other copies. From these adversarial examples, the investigator wants to identify the source $\mathcal{M}_i$. 
To address this problem, we propose a two-stage separate-and-trace framework. The model separation stage generates multiple copies of a model for a same classification task. This process injects unique characteristics into each copy so that adversarial examples generated have distinct and traceable features. We give a parallel structure which embeds a ``tracer'' in each copy, and a noise-sensitive training loss to achieve this goal. The tracing stage takes in adversarial examples and a few candidate models, and identifies the likely source. Based on the unique features induced by the noise-sensitive loss function, we could effectively trace the potential adversarial copy by considering the output logits from each tracer. 
Empirical results show that it is possible to trace the origin of the adversarial example and the mechanism can be applied to a wide range of architectures and datasets.
\end{abstract}

\section{Introduction}
Deep learning models are vulnerable to adversarial attacks. By introducing specific perturbations on input samples, the network model could be misled to give wrong predictions even when the perturbed sample looks visually close to the clean image \cite{szegedy2014intriguing,goodfellow2014explaining,moosavi2016deepfool,carlini2017towards}. There are many existing works on defending against such attacks
\cite{kurakin2016adversarial,meng2017magnet,gu2014towards,hinton2015distilling}. 
Unfortunately, although current defenses could mitigate the attack to some extent, the threat is still far from being completely eliminated.
In this paper, we look into the forensic aspect: from the adversarial examples, can we determine which model the adversarial examples were derived from? Techniques derived could aid forensic investigation of attack incidents and provide deterrence to future attacks.

We consider a buyers-seller setting~\cite{zhang2021mitigating}, which is similar to the buyers-seller setting in digital rights protection~\cite{memon2001buyer}.

\begin{figure}[]
 \includegraphics[width=1\linewidth]{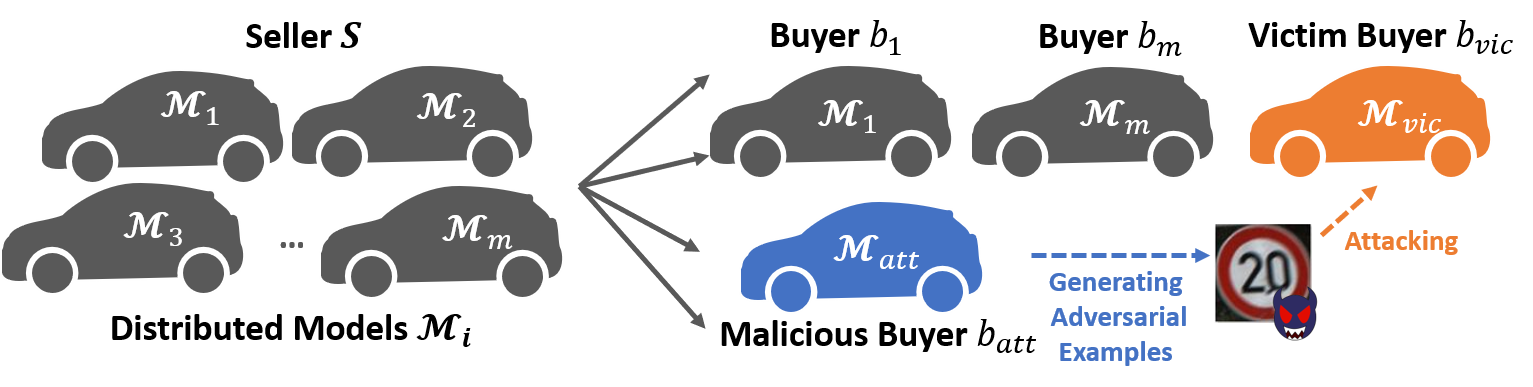}
\caption{\textit{Buyers-seller setting.} The seller has multiple models $\mathcal{M}_i, i\in[1,m]$ that are to be distributed to different buyers. A malicious buyer $b_{att}$ attempts to attack the victim buyer $b_{vic}$ by generating the adversarial examples with his own model $\mathcal{M}_{att}$.}
\label{TeaserFig}
\vspace{-10pt}
\end{figure}

\paragraph{\textit{Buyers-seller Setting.}} Under this setting, the seller $\textbf{\textit{S}}$ distributes $m$  classification models $\mathcal{M}_i, i\in[1,m]$  to different buyers $b_i$'s as shown in Fig. \ref{TeaserFig}. These models are trained for a same classification task using a same training dataset. The models are made accessible to the buyer as black boxes,  for instance,  the models could be embedded in hardware such as FPGA and ASIC,  or are provided in a  Machine Learning as a Service (MLaaS) platform. Hence, the buyer only has black-box access, which means that he can only query the model for the hard label. In addition, we assume that the buyers do not know the training datasets. The seller has full knowledge and thus has white-box access to all the distributed models. 

\paragraph{\textit{Attack and Traceability.}}  A malicious buyer wants to attack other victim buyers. The malicious buyer does not have direct access to other models and thus generates the examples from its own model and then deploys the found examples. For example, the malicious buyer might generate an adversarial example of a road sign using its self-driving vehicle, and then physically defaces the road sign to trick passing vehicles.  Now, as forensic investigators who have obtained the defaced road sign, we want to understand how the adversarial example is generated and trace the models used in generating the example. 

\paragraph{\textit{Proposed Framework.}} There are two stages in our solution: {\em model separation} and {\em origin tracing}.  During the model separation stage, given a classification task, we want to generate multiple models that have high accuracy on the classification task and yet are sufficiently different for tracing. In other words, we want to proactively enhance differences among the models in order to facilitate tracing.  To achieve that, we propose a parallel network structure that pairs a unique tracer with the original classification model.
The role of the tracer is to modify the output, so as to induce the attacker to adversarial examples with unique features. We give a noise-sensitive training loss for the tracer.   

During the tracing stage, given $m$ different classification models $\mathcal{M}_i, i\in[1,m]$ and the found adversarial example, we want to determine which model is most likely used in generating the adversarial examples.  This is achieved by exploiting the different tracers that are earlier embedded into the parallel models. Our proposed method compares the output logits (the output of the network before softmax) of those tracers to identify the source.


 
In a certain sense, traceability is similar to neural network watermarking and can be viewed as a stronger form of watermarking. Neural network watermarking schemes \cite{boenisch2020survey} attempt to generate multiple models so that an investigator can trace the source of a modified copy.  In traceability, the investigator can trace the source based on the generated adversarial examples.

\paragraph{\textit{Contributions.}}
\begin{enumerate}[leftmargin=*]
\item We point out a new aspect in defending against adversarial attacks, that is, tracing the origin of adversarial samples among multiple classifiers. Techniques derived would aid forensic investigation of attack incidents and provide deterrence to future attacks.


\item We propose a framework to achieve traceability in the buyers-seller setting. The framework consists of two stages: a model separation stage, and a tracing stage.  The model separation stage generates multiple ``well-separated'' models and this is achieved by a parallel network structure that pairs a tracer with the classifier. The tracing mechanism exploits the characteristics of the paired tracers to decide the origin of the given adversarial examples.

\item We investigate the effectiveness of the separation and the subsequent tracing. Experimental studies show that the proposed mechanism can effectively trace to the source.  For example, the tracing accuracy achieves more than 97\% when applying to ``ResNet18-CIFAR10'' task. We also observe a clear separation of the source tracer's  logits distribution, from the non-source's logits distribution (e.g. Fig. 5a-5c). 

\end{enumerate}

\section{Related Work}\label{RelatedWork}
In this paper, we adopt black-box settings where the adversary can only query the model and get the hard label (final decision) of the output. Many existing attacks assume white-box settings. Attack such as FGSM \cite{goodfellow2014explaining}, PGD \cite{kurakin2016adversarial}, JSMA \cite{papernot2016limitations}, DeepFool \cite{moosavi2016deepfool}, CW \cite{carlini2017towards} and EAD \cite{chen2018ead} usually directly rely on the gradient information provided by the victim model. As the detailed information of the model is hidden in black-box settings, black-box attacks are often considered more difficult and there are fewer works. Chen \textit{et. al.} introduced a black-box attack called Zeroth Order Optimization (ZOO) \cite{chen2017zoo}. ZOO can approximate the gradients of the objective function with finite-difference numerical estimates by only querying the network model. Thus the approximated gradient is utilized to generate the adversarial examples. Guo \textit{et. al.} proposed a simple black-box adversarial attack called ``SimBA'' \cite{guo2019simple} to generate adversarial examples with a set of orthogonal vectors. By testing the output logits with the added chosen vector, the optimization direction can be effectively found. Brendel \textit{et. al.} developed a decision-based adversarial attack which is known as ``Boundary attack'' \cite{brendel2018decision}, it worked by iteratively perturbing another initial image that belongs to a different label toward the decision boundaries between the original label and the adjacent label. By querying the model with enough perturbed images, the boundary as well as the perturbation can be found thus generating the adversarial examples. Chen \textit{et. al.} proposed another decision based attack named hop-skip-jump attack (HSJA) \cite{chen2020hopskipjumpattack} recently. By only utilizing the binary information at the decision boundary and the Monte-Carlo estimation, the gradient direction of the network can be found so as to realize the adversarial examples generation. Based on \cite{chen2020hopskipjumpattack}, Li \textit{et. al.} \cite{li2020qeba} proposed a query-efficient boundary-based black-box attack named QEBA which estimate the gradient of the boundary in several transformed space and effectively reduce the query numbers in generating the adversarial examples. Maho \textit{et. al.} \cite{maho2021surfree} proposed a surrogate-free black-box attack which do not estimate the gradient but searching the boundary based on polar coordinates, compared with \cite{chen2020hopskipjumpattack} and \cite{li2020qeba}, \cite{maho2021surfree} achieves less distortion with less query numbers.

\section{Proposed Framework}\label{ProposedFramework}
\begin{figure*}
    \begin{minipage}[t]{1\linewidth}
        \centering{
            \includegraphics[width=0.9\linewidth]{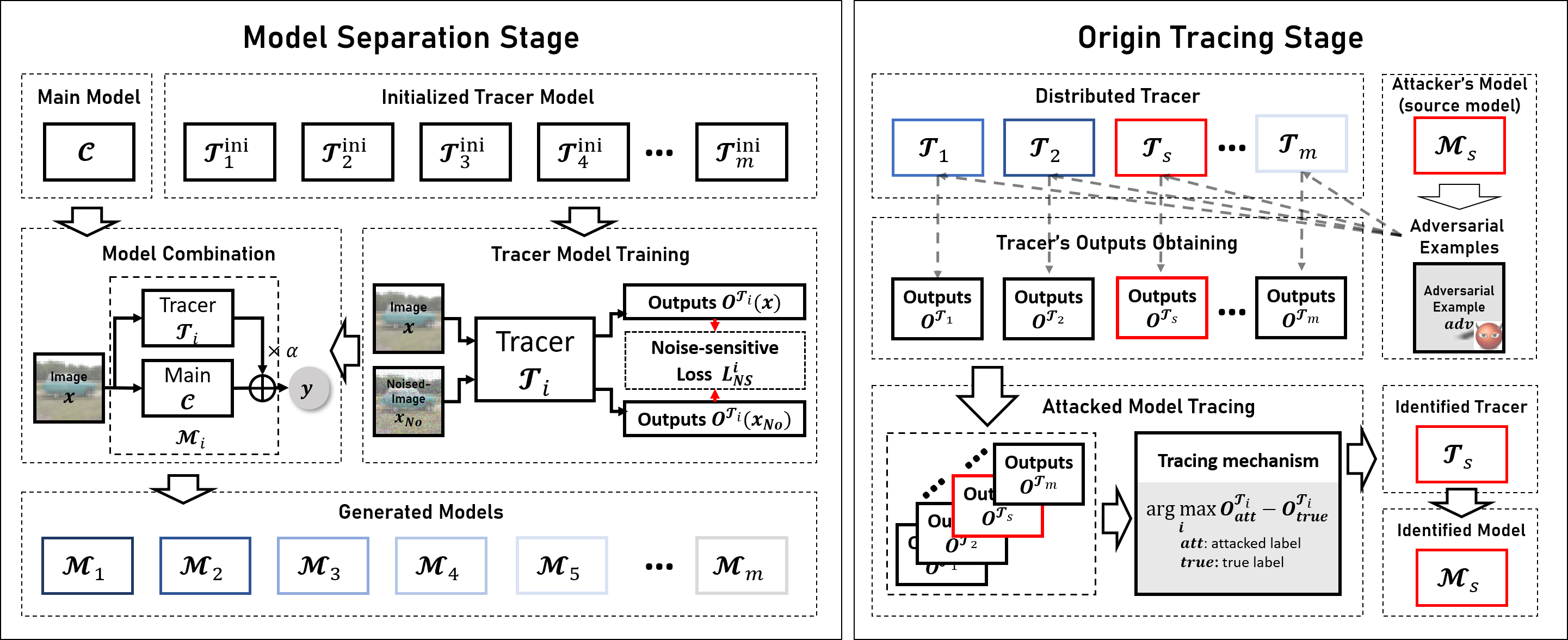}
        }
    \end{minipage}%
\caption{The framework of the proposed method. The left part of the framework indicates the separation process of the seller's distributed models $\mathcal{M}_i, i\in[1,m]$. The right part of the framework illustrates the origin tracing process.}
\label{FrameworkFig}
\vspace{-10pt}
\end{figure*}
\subsection{Main Idea}\label{MainIdea}
We design a framework that contains two stages: model separation and origin tracing. 

During the model separation stage, we want to generate multiple models which are sufficiently different under adversarial attack while remaining highly accurate on the classification task. Our main idea is a parallel network structure which pairs a unique tracer with the original classifier. The specific structure will be illustrated in Section \ref{ModelTraining}. 

As for origin tracing, we exploit unique characteristics of different tracers in the parallel structure, which can be observed in the tracers' logits.  Hence, our tracing process is conducted by feeding the adversarial examples into the tracers and analyzing their output.

The whole framework of the proposed scheme is shown in Fig. \ref{FrameworkFig}. As illustrated in Fig.\ref{FrameworkFig}, each distributed model $\mathcal{M}_i$ consists of a tracer $\mathcal{T}_i$ and the original classification model $\mathcal{C}$, and the tracer is trained with a proposed noise-sensitive loss  $\mathcal{L}_{NS}$.   During the tracing stage, the adversarial examples are fed into each $\mathcal{T}_i$ and the outputs are analyzed to identify the origin.






\subsection{Model Separation}\label{ModelTraining}

\begin{figure*}[t]
\begin{center}
    \subfloat[Parallel network structure.]{
    \label{Fig_PN}
        {\centering\includegraphics[width=0.3\linewidth]{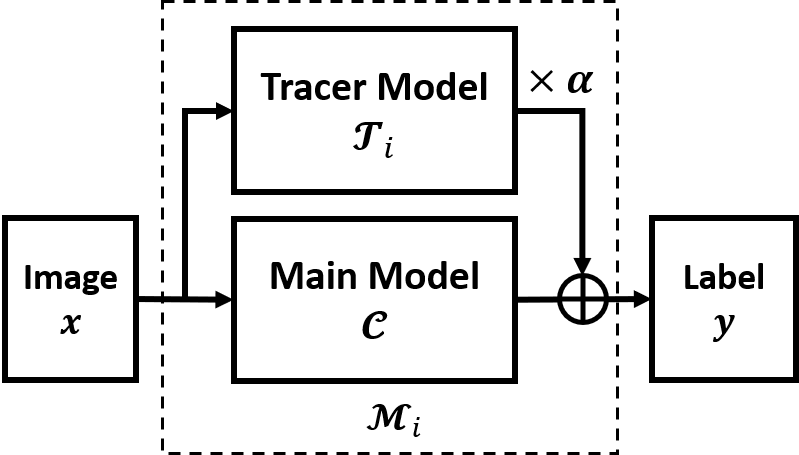}
        }
    }
    \subfloat[The architecture of tracer.]{
    \label{Fig_tracer}
        {\centering\includegraphics[width=0.25\linewidth]{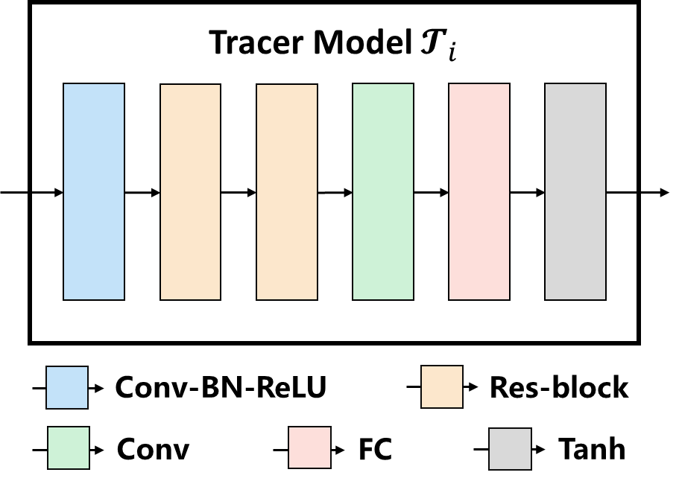}
        }
    }
    \subfloat[Differences in logits.]{
    \label{Fig_DL}
        {\centering\includegraphics[width=0.25\linewidth]{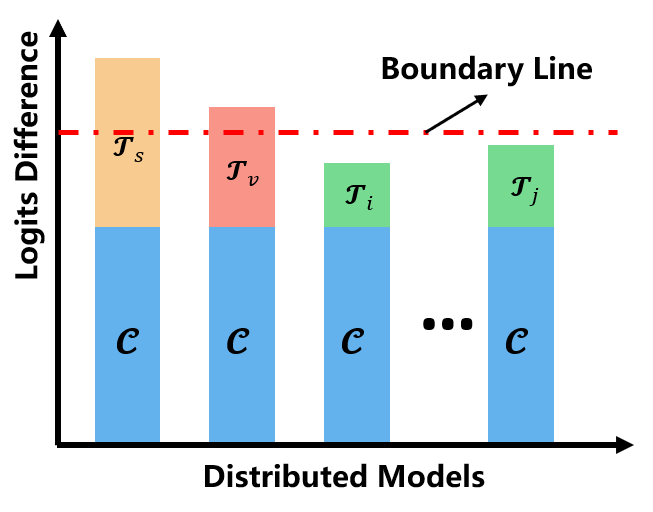}
        }
    }
    
\end{center}
\vspace{-5pt}
\caption{The specific network design in model separation.}
\label{ModelSeparation}
\vspace{-10pt}
\end{figure*}

We design a parallel network structure to generate the distributed models $\mathcal{M}_i, i\in[1,m]$, which contains a tracer model $\mathcal{T}_i$ and a main model $\mathcal{C}$, as shown in Fig. \ref{Fig_PN}. $\mathcal{T}_i$ is used for injecting unique features and setting traps for the attacker. $\mathcal{C}$ is the network trained for the original task. The final results are determined by both $\mathcal{C}$ and $\mathcal{T}_i$ with a weight parameter $\alpha$. In each distributed model, $\mathcal{C}$ is fixed and only $\mathcal{T}_i$ is different.

The specific structure of $\mathcal{T}_i$ is shown in Fig. \ref{Fig_tracer}, it is linearly cascaded with one ``SingleConv'' block (Conv-BN-ReLU), two ``Res-block'' \cite{he2016deep}, one ``Conv'' block, one full connection block and one ``Tanh'' activation layer. The training process of $\mathcal{T}_i$ can be described as:


1) Given the training dataset\footnote{The training dataset for $\mathcal{T}_i$ only contains 1000 random sampled images from the dataset of the original classification task} and tracer $\mathcal{T}_i$, we first initialize $\mathcal{T}_i$ with random parameters.

2) For each training epoch, we add random noise $No$ \footnote{$No$ follows a uniform distribution over [0, 0.03)} on the input image $x$ to generate the noised image $x_{No}$. 

3) Then we feed both $x$ and $x_{No}$ into $\mathcal{T}_i$ and get the outputs $O_x$ and $O_{x_{No}}$. We attempt to make $\mathcal{T}_i$ sensitive to noise, so $O_x$ and $O_{x_{No}}$ should be as different as possible. The loss function of $\mathcal{T}_i$ can be written as:
\begin{equation}
\label{eq_L}
\mathcal{L}_{NS}=\frac{\left|O_{x} \circ O_{x_{No}}\right|}{\lVert O_{x}\rVert_{{2}}\lVert O_{x_{No}}\rVert_{{2}}} = \frac{\left|\mathcal{T}_i(\theta_{\mathcal{T}_i},x)\circ \mathcal{T}_i(\theta_{\mathcal{T}_i},x_{No})\right|}{\lVert  \mathcal{T}_i(\theta_{\mathcal{T}_i},x)\rVert_{{2}}\lVert \mathcal{T}_i(\theta_{\mathcal{T}_i},x_{No})\rVert_{{2}}}
\end{equation}
where $\circ$ represents the Hadamard product. $\theta_{\mathcal{T}_i}$ indicates the parameters of $\mathcal{T}_i$. 

Each distributed $\mathcal{T}_i$ for different buyers is generated by randomly initializing and then training. We believed the randomness in initialization is enough to guarantee the difference from different $\mathcal{T}_i$. It should be noted that when producing a new distributed copy, we only have to train one new tracer without setting more constraints on former tracers. So such a separation method can be applied to multiple distributed models independently.

As for $\mathcal{C}$, it is trained in a normal way which utilizes the whole training dataset and cross-entropy loss. For the main classification task, $\mathcal{C}$ only has to be trained once. Besides, the training of $\mathcal{C}$ is independent of the training of $\mathcal{T}_i$. After training $\mathcal{C}$, we could get a high accuracy classification model. The final distributed model $\mathcal{M}_i$ is parallel combined with $\mathcal{C}$ and $\mathcal{T}_i$. The specific workflow of $\mathcal{M}_i$ can be described as:

For input image $x$, $\mathcal{T}_i$ and $\mathcal{C}$ both receive the same $x$ and output two different vectors $O^{\mathcal{T}_i}$ and $O^{\mathcal{C}}$ respectively. $O^{\mathcal{T}_i}$ and $O^{\mathcal{C}}$ have the same size and will be further added in a weighted way to generate the final outputs $O^F$, as shown in Eq. \ref{eq_OF}.
\begin{equation}
\label{eq_OF}
O^{F} = O^{\mathcal{C}} + \alpha \times O^{\mathcal{T}_i}
\end{equation}
where $\alpha$ is the weight parameter. It is worth noting that for the output of $\mathcal{C}$, we use the normalization form of it, which can be formulated as:
\begin{equation}
\label{eq_OC}
O^{\mathcal{C}} = \frac{\mathcal{C}(x) - \min(\mathcal{C}( x))}{\max(\mathcal{C}(x))-\min(\mathcal{C}(x))}
\end{equation}
where $x$ indicates the input image, $\max$ and $\min$ indicate the maximum value and minimum value respectively.

By utilizing the aforementioned model separation method, two properties are well satisfied: (I) The attack could be tricked to focus more on $\mathcal{T}_i$ than $\mathcal{C}$. Since after the training, $\mathcal{T}_i$ will be sensitive to random noise. Therefore, the output of $\mathcal{T}_i$ is easy to be changed by adding noise. Compared with $\mathcal{C}$, the boundary of $\mathcal{T}_i$ is more likely to be estimated and $\mathcal{T}_i$ is more likely to be attacked. Thus, the attacker will fall into the trap of $\mathcal{T}_i$ and the generated adversarial perturbations will bring the feature of the source $\mathcal{T}_i$. (II) Based on random initialization, each distributed $\mathcal{T}_i$ will correspond to different adversarial perturbations. 
This property helps us in tracing, since the source $\mathcal{T}_{s}$ which generates adversarial examples will output unique responses compared with other $\mathcal{T}_i, i\neq s$ when feeding the generated adversarial examples, as shown in Fig. \ref{Fig_DL}.

\subsection{Tracing the Origin}\label{tracerTraining}
The tracing process is conducted by two related components:
\begin{itemize}
    \item The first component keeps white-box copies for each of the $m$ distributed copies~\footnote{This setting is reasonable because when an adversarial attack appeared, the model seller who has all the details of the distributed network takes responsible to trace the attacker. }. This component allows us to obtain the output logits of each tracer on an input $x$. 
    \item The second component is an output logits-based mechanism. It gives a decision on which copy $i$ is the most likely one to generate the adversarial example.
\end{itemize}

The specific tracing process can be described as follows:

1) Given an appeared adversarial examples denoted as $x_{att}$, we feed the adversarial example into all $\mathcal{T}_i, i\in[1,m]$ and obtain the output logits of them, noted as $O^{\mathcal{T}_i}, i \in[1,m]$.

2) Then we extract two values that are corresponding to the attacked label and true label in each $O^{\mathcal{T}_i}$, denoted as $O^{\mathcal{T}_i}_{att}$ and $O^{\mathcal{T}_i}_{true}$ respectively. \footnote{Attacked label can be easily determined by the output logits and the true label can be tagged by the model owner. If this sample cannot be accurately tagged by the owner, then this sample is not regarded as an adversarial example.} 

3) The source model can be determined by:
\begin{equation}
\label{eq_am}
s=\underset{i, i \in[1, m]}{\arg \max }( O^{\mathcal{T}_i}_{att} - O^{\mathcal{T}_i}_{true} )
\end{equation}
To simplify the description, we denote the difference of output logits ($O^{\mathcal{T}_i}_{att} - O^{\mathcal{T}_i}_{true}$) as DOL. The tracer corresponded to the largest DOL is regarded as the source model. The reason is as follows:

Since the perturbation are highly related to $\mathcal{T}_i$, when feeding the same adversarial example, the outputs of $\mathcal{T}_i$ and $\mathcal{T}_j$ ($i\neq j$)  will be certainly different. For source model $\mathcal{T}_{s}$ where the adversarial examples are generated from, $O^{\mathcal{T}_s}$ is likely to render a large value on the adversarial label and a small value on the ground-truth label. Since the weight of $O^{\mathcal{T}_s}$ in the final $O^{\mathcal{F}_s}$ is small, so in order to achieve adversarial attack, $O^{\mathcal{T}_s}$ will be modified as much as possible. Thus DOL of $\mathcal{T}_{s}$ should be large. But for victim model $\mathcal{T}_{v}$, the DOL will be small. Therefore, according to the value of DOL, we can trace the origin of the adversarial example.

\section{Experimental Results}\label{EXP}
\subsection{Implementation Details}\label{Implementation}
In order to show the effectiveness of the proposed framework, we perform the experiments on two network architecture (ResNet18 \cite{he2016deep} and VGG16 \cite{simonyan2014very}) with two small image datasets (CIFAR10 \cite{krizhevsky2009learning} of 10 classes and GTSRB \cite{houben2013detection} of 43 classes) and two deeper network architecture (ResNet50 and VGG19) with one big image dataset (mini-ImageNet \cite{ravi2016optimization} of 100 classes). The main classifier $\mathcal{C}$ in experiments is trained for 200 epochs. All the model training is implemented by PyTorch and executed on NVIDIA RTX 2080ti. For gradient descent, Adam \cite{kingma2014adam:} with learning rate of 1e-4 is applied as the optimization method. 
\subsection{The Classification Accuracy of The Proposed Architecture}\label{Alpha_Value}
The most influenced parameter for the classification accuracy is the weight parameter $\alpha$. $\alpha$ determines the participation ratio of $\mathcal{T}_i$ in final outputs. To investigate the influence of $\alpha$, we change the value of $\alpha$ from 0 (baseline) to 0.2 and record the corresponding classification accuracy of each task, the results are shown in Table \ref{ModelAccuracy}.
\begin{table}[h]

\centering
\scalebox{0.78}{
\begin{tabular}{c|cc|cc|cc}
\toprule[2pt]
\multirow{2}{*}{$\alpha$} & \multicolumn{2}{c|}{CIFAR10}     & \multicolumn{2}{c|}{GTSRB}  & \multicolumn{2}{c}{Mini-ImageNet} \\
        & ResNet18     & VGG16     & ResNet18     & VGG16  & ResNet50     & VGG19 \\
\midrule
0      & $94.30\%$ & $93.68\%$ & $96.19\%$ & $97.59\%$ & $73.12\%$ & $75.79\%$\\
\midrule
0.05       & $94.24\%$ & $93.64\%$ & $96.14\%$ & $97.52\%$ & $72.32\%$ & $75.04\%$\\
\midrule
0.1        & $94.24\%$ & $93.63\%$ & $96.07\%$ & $97.36\%$ & $71.88\%$ & $74.96\%$\\
\midrule
0.15       & $94.07\%$ & $93.63\%$ & $95.72\%$ & $96.84\%$ & $70.50\%$ & $73.75\%$\\
\midrule
0.2       & $93.95\%$ & $93.57\%$ & $95.09\%$ & $95.52\%$ & $68.14\%$ & $71.75\%$\\
\bottomrule[2pt]
\end{tabular}}

\caption{The classification accuracy with different $\alpha$.}
\label{ModelAccuracy}
\end{table}

It can be seen from Table \ref{ModelAccuracy} that for CIFAR10 and GTSRB, the growth of $\alpha$ will seldom decrease the accuracy of the classification task. Compared with the baseline ($\alpha=0$), the small value of $\alpha$ will keep the accuracy at the same level as the baseline. But for mini-ImageNet, the accuracy decreases more as $\alpha$ increases, we believe it is due to the complexity of the classification task. But even though, the decrease rate is still within 3\% when $\alpha$ is not larger than 0.15.


\subsection{Traceability of different black-box attack}\label{Traceability}
It should be noted that the change of $\alpha$ will not only influence the accuracy but also affect the process of black-box adversarial attack. Therefore, in order to explore the influence of $\alpha$, the following experiments will be conducted with $\alpha=0.05, 0.1$ and $0.15$. 

\textbf{Setup and Code.}\ \ 
To verify the traceability of the proposed mechanism, we conduct experiments on two distributed models. We set one model as the source model $\mathcal{M}_s$ to perform the adversarial attack and set the other model as the victim model $\mathcal{M}_v$. The goal is to test whether the proposed scheme can effectively trace the source model from the generated adversarial examples. The black-box attack we choose is Boundary \cite{brendel2018decision}, HSJA \cite{chen2020hopskipjumpattack}, QEBA \cite{li2020qeba} and SurFree \cite{maho2021surfree}. For Boundary \cite{brendel2018decision} and HSJA \cite{chen2020hopskipjumpattack}, we use Adversarial Robustness Toolbox (ART) \cite{nicolae2018adversarial} platform to conduct the experiments. For QEBA \cite{li2020qeba} and SurFree \cite{maho2021surfree}, we pull implementations from their respective GitHub repositories \footnote{QEBA:https://github.com/AI-secure/QEBA} \footnote{SurFree:https://github.com/t-maho/SurFree} with default parameters. For each $\alpha$, each network architecture, each dataset and each attack, we generate 1000 successful attacked adversarial examples of $\mathcal{M}_s$ and conduct the tracing experiment.

\textbf{Evaluation Metrics.}\ \ 
Traceability is evaluated by tracing accuracy, which is calculated by:
\begin{equation}
\label{eq_Acc}
\text { Acc }=\frac{N_{\text {correct }}}{N_{\text {All }}}
\end{equation}
where $N_{\text {correct }}$ indicates the number of correct-tracing samples and  $N_{\text {All }}$ indicates the total number of samples, which is set as 1000 in the experiments. 

\begin{table*}[h]
\centering

\vspace{-5pt}
\scalebox{0.81}{
\begin{tabular}{cc|ccc|ccc|ccc|ccc}
\toprule[2pt]
\multicolumn{2}{c|}{Attack}                 & \multicolumn{3}{c|}{Boundary} & \multicolumn{3}{c|}{HSJA} & \multicolumn{3}{c|}{QEBA} & \multicolumn{3}{c}{SurFree} \\\midrule
\multicolumn{2}{c|}{alpha}                  & 0.05     & 0.1     & 0.15    & 0.05   & 0.1    & 0.15   & 0.05   & 0.1    & 0.15   & 0.05    & 0.1     & 0.15    \\
\midrule
\multirow{2}{*}{CIFAR10}        & ResNet18 & 98.1\%     & 98.9 \%   & 99.2 \%   & 98.2\%   & 99.1\%   & 99.3\%   & 99.6\%   & 99.7\%   & 99.7\%   & 94.5\%    & 95.7\%    & 97.9\%    \\
                                & VGG16    & 92.1\%     & 95.6 \%   & 98.2\%    & 92.3 \%  & 96.4 \%  & 97.9 \%  & 92.6 \%  & 96.6\%   & 99.2\%   & 64.2 \%   & 82.1 \%   & 87.8 \%   \\
                                \midrule
\multirow{2}{*}{GTSRB}          & ResNet18 & 97.6\%     & 97.6 \%   & 98.9  \%  & 97.6 \%  & 97.7 \%  & 98.7 \%  & 97.6\%   & 97.7\%   & 99.6 \%  & 89.8 \%   & 95.7 \%   & 96.8\%    \\
                                & VGG16    & 94.1 \%    & 96.8 \%   & 97.6 \%   & 95.5 \%  & 97.3 \%  & 98.3\%   & 86.3\%   & 92.6\%   & 95.0 \%  & 89.7\%    & 95.7\%    & 96.8\%    \\
                                \midrule
\multirow{2}{*}{mini\_ImageNet} & ResNet50 & 96.2\%     & 96.4 \%   & 98.7 \%   & 94.5\%   & 95.5 \%  & 97.5 \%  & 91.7\%   & 93.8\%   & 95.4 \%  & 82.1 \%   & 87.3 \%   & 90.5\%    \\
                                & VGG19    & 89.4 \%    & 94.7  \%  & 98.2\%    & 93.4  \% & 95.1 \%  & 95.4\%   & 89.5\%   & 90.4\%   & 90.8 \%  & 75.7 \%   & 88.7 \%   & 88.8\%   \\
\bottomrule[2pt]
\end{tabular}
}
\caption{The trace accuracy of different attacks.}
\label{TraceAccuracy}
\vspace{-10pt}
\end{table*}
The tracing performance of different attacks with different settings is shown in Table \ref{TraceAccuracy}. It can be seen that when applying ResNet-based architecture as the backbone of $\mathcal{C}$, the tracing accuracy is higher than 90\%. Especially for $\alpha=0.15$, most of the tracing accuracy is higher than 96\%, which indicates the effectiveness of the proposed mechanism. Besides, for a different level of classification task and different attacking methods, the tracing accuracy can stay at a high level, which shows the great adaptability of the proposed scheme. 

\begin{figure}[H]
\vspace{-15pt}
\begin{center}
    \subfloat[The results of Boundary.]{
    \label{Fig_HSJA_OD}
        {\centering\includegraphics[width=0.45\linewidth]{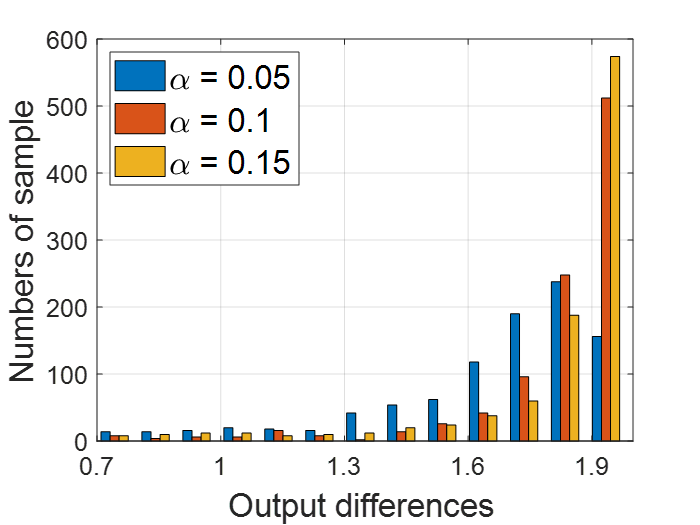}
        }
    }
    \subfloat[The results of HSJA.]{
    \label{Fig_QEBA_OD}
        {\centering\includegraphics[width=0.45\linewidth]{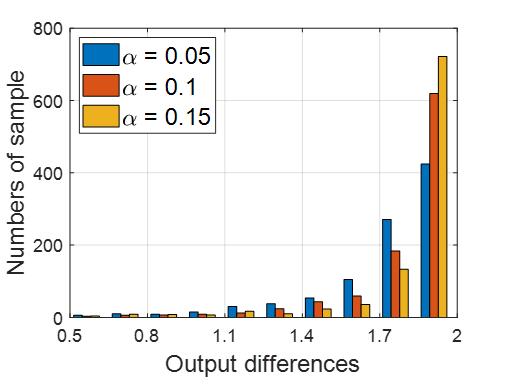}
        }
    }
    \\[-10pt]
    \subfloat[The results of QEBA.]{
    \label{Fig_HSJA_OD}
        {\centering\includegraphics[width=0.45\linewidth]{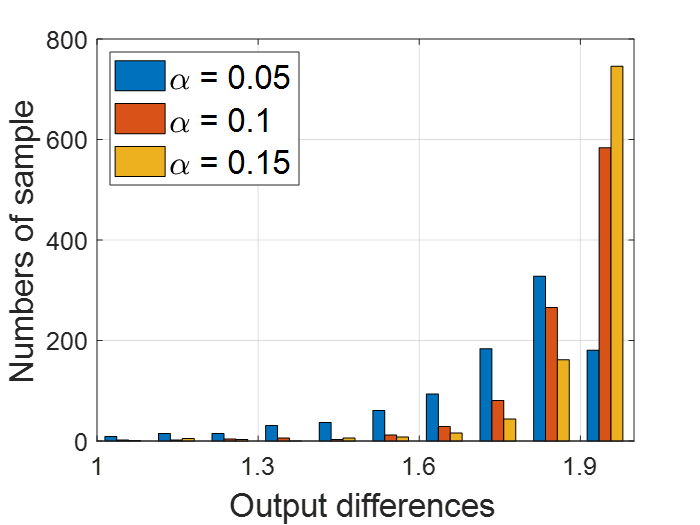}
        }
    }
    \subfloat[The results of SurFree.]{
    \label{Fig_QEBA_OD}
        {\centering\includegraphics[width=0.45\linewidth]{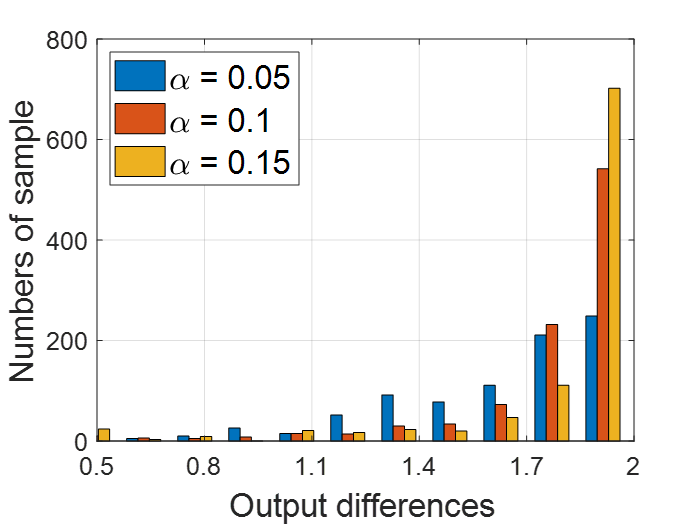}
        }
    }
\end{center}
\vspace{-5pt}
\caption{The distributions of output differences with different black-box attacks.}
\label{FIG_OD}
\end{figure}

\textbf{The influence of $\alpha$.}\ \ 
We can see from Table \ref{TraceAccuracy} that the tracing accuracy increases with the increase of $\alpha$. We conclude the reason as: $\alpha$ determines the participation rate of tracer $\mathcal{T}_i$ in final output logits, the larger $\alpha$ will make the final decision boundary rely more on $\mathcal{T}$. Therefore, when $\alpha$ gets larger, making DOL of $\mathcal{T}$ larger would be a better choice to realize the adversarial attack. The bigger DOL of $\mathcal{T}$ will certainly lead to better tracing performance. 
To verify the correctness of the explanation, we show the distribution of DOL for task ``ResNet18-CIFAR10'' with different attacks in Fig. \ref{FIG_OD}. We first generate 1000 adversarial examples of model $\mathcal{M}_{i}$ for each $\alpha$ ($0.05$,$0.1$,$0.15$) with Boundary, HSJA, QEBA and SurFree attack, then we record the DOLs of $\mathcal{T}_{i}$. The distribution of DOLs are shown in Fig. \ref{FIG_OD}. 

It can be seen that compared with $\alpha=0.05$ and $\alpha=0.1$, the DOL of $\alpha=0.15$ concentrate more on larger values, which indicates that the larger $\alpha$ will result to larger DOL. 

\textbf{The influence of network architecture.}\ \ 
The tracing results vary with different networks and different datasets. With the same dataset, the tracing accuracy of ResNet18 will be higher than that of VGG16. We attribute the reason to the complexity of the model architecture. According to \cite{su2018robustness}, compared with ResNet, the structure of VGG is less robust, so VGG-based $\mathcal{C}$ might be easier to be adversarial attacked. Therefore, once $\mathcal{C}$ is attacked, there is a certain probability that $\mathcal{T}_i$ is not attacked as we expected, so DOL of $\mathcal{T}_i$ will not produce the expected features for tracing. Fortunately, the network architecture can be designed by us, so in practice, choosing a robust architecture would be better for tracing. 

\textbf{The influence of classification task.} \ \
In our experiments, we test the classification task with different classes. It can be seen that with the increase of classification task complexity, traceability performance decreases slightly. But in most cases, when $\alpha = 0.15 $, the traceability ability can still reach more than 90\%.

\textbf{The influence of black-box attack.}\ \ 
The mechanism of the black-box attack greatly influences the tracing performance. For Boundary attack\cite{brendel2018decision}, HSJA\cite{chen2020hopskipjumpattack} and QEBA\cite{li2020qeba}, the tracing accuracy shows similar results, but for SurFree \cite{maho2021surfree}, the tracing accuracy will be worse than that of the other attacks. The reason is that Boundary attack, HSJA\cite{chen2020hopskipjumpattack}, QEBA\cite{li2020qeba} are gradient-estimation-based attacks, which tries to use random noise to estimate the gradient of the network and further attack along the gradient. Since the gradient is highly related to $\mathcal{T}_i$, such attacks are more likely to be trapped by $\mathcal{T}_i$. But SurFree\cite{maho2021surfree} is attacking based on geometric characteristics of the boundary, which may ignore the trap of $\mathcal{T}_i$ especially when $\alpha$ is small. So compared with Boundary attack\cite{brendel2018decision}, HSJA\cite{chen2020hopskipjumpattack} and QEBA\cite{li2020qeba}, the proposed mechanism may get worse performance when facing SurFree\cite{maho2021surfree} attack.

\begin{figure*}[t]
\begin{center}
    \subfloat[The distribution of CIFAR10.]{
    \label{Fig_D_CIFAR10}
        {\centering\includegraphics[width=0.3\linewidth]{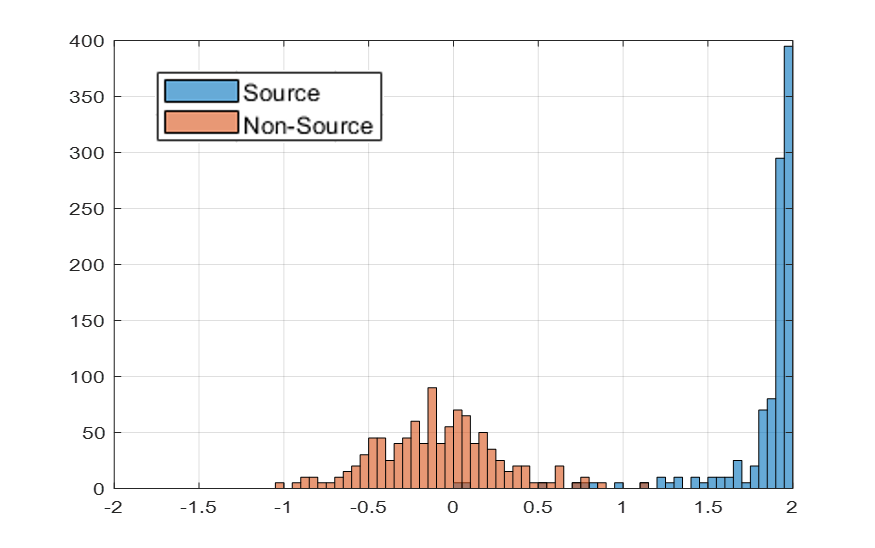}
        }
    }
    \subfloat[The distribution of GTSRB.]{
    \label{Fig_D_GTSRB}
        {\centering\includegraphics[width=0.3\linewidth]{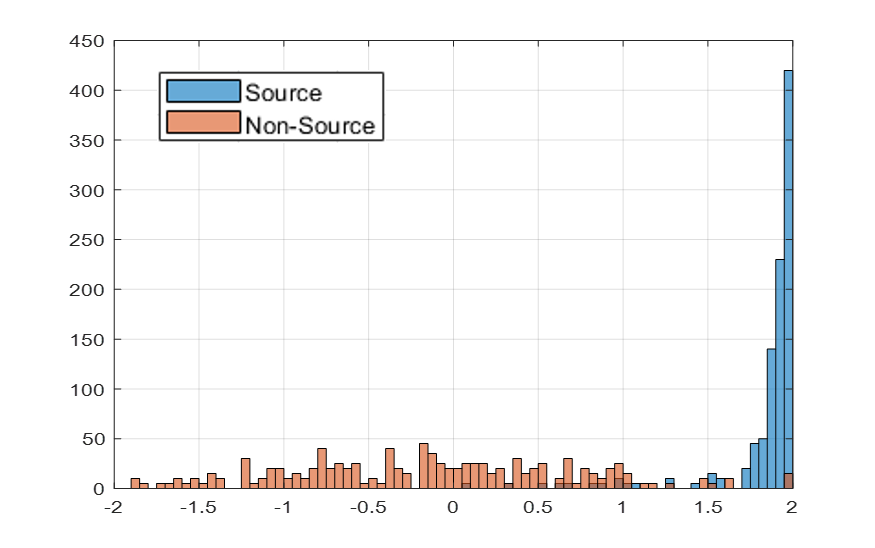}
        }
    }
    \subfloat[The distribution of mini-ImageNet.]{
    \label{Fig_D_MIN}
        {\centering\includegraphics[width=0.3\linewidth]{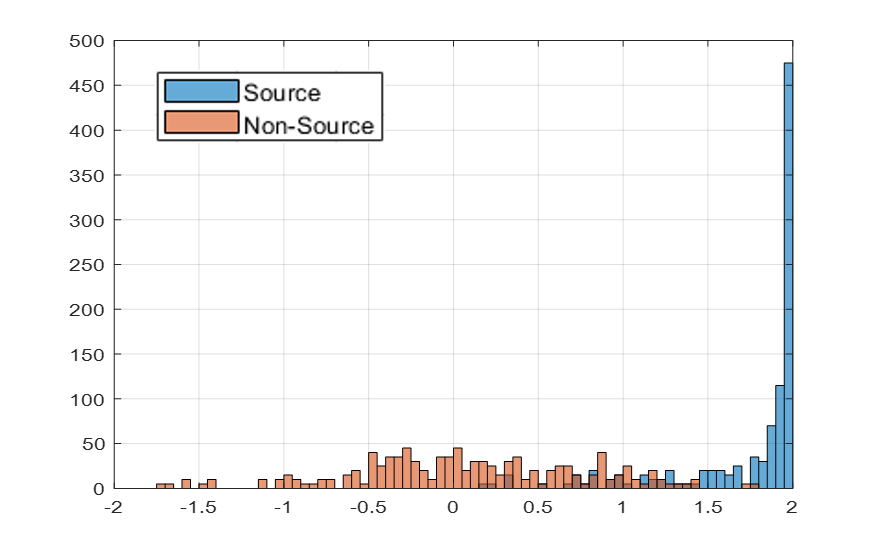}
        }
    }
    \\[-10pt]
    \subfloat[The tracing results of CIFAR10.]{
    \label{Fig_CIFAR10}
        {\centering\includegraphics[width=0.3\linewidth]{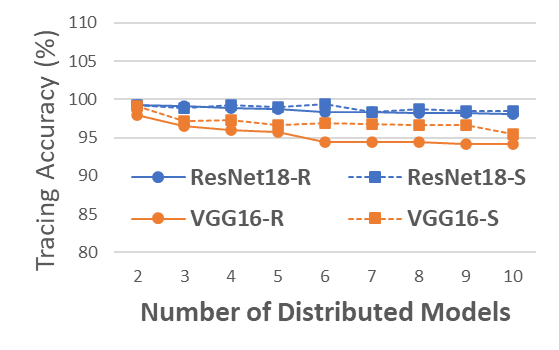}
        }
    }
    \subfloat[The tracing results of GTSRB.]{
    \label{Fig_GTSRB}
        {\centering\includegraphics[width=0.3\linewidth]{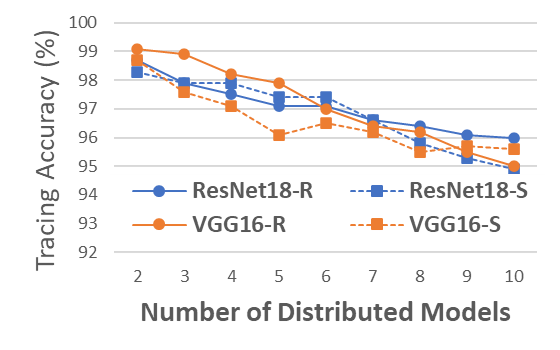}
        }
    }
    \subfloat[The tracing results of mini-ImageNet.]{
    \label{Fig_MIN}
        {\centering\includegraphics[width=0.3\linewidth]{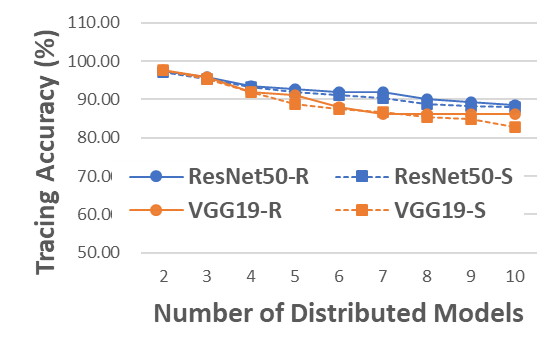}
        }
    }
\end{center}
\caption{The distribution of DOL with HSJA and ResNet backbone and tracing performance of multiple branches.}
\label{FIG_NBranch}

\end{figure*}

\subsection{The influence of distributed copy numbers}
In this section, we will discuss the traceability of the algorithm in multiple distributed copies. When training tracer $\mathcal{T}_i$, the parameter is randomly initialized and each $\mathcal{T}_i$ is trained independently. So the distribution of DOL corresponding to any two branches should follow independent and identically distribution. Therefore, the traceability results of multiple copies could be calculated from the results of two copies. In order to verify the correctness, we perform the following experiments. 

For experiment verification, we trained 10 different $\mathcal{T}_i$ first, then we randomly choose one $\mathcal{M}_s$ as the source model to generate the adversarial examples. We record the tracing performance on the $n, n\in[2,10]$ models. 

To estimate the tracing results for $n, n\in[2,10]$ models, we utilize the Monte-Carlo sampling method in the distribution of two models' DOL. The specific procedure is described as:

1). We randomly choose one source model $\mathcal{M}_{s}$ and one other victim model $\mathcal{M}_{v}$ as the fundamental models, then we perform the black-box attack on $\mathcal{M}_{s}$ with 1000 different images and record the DOL of $\mathcal{T}_{s}$ and $\mathcal{T}_{v}$.

2). We draw the distribution of DOL corresponding to $\mathcal{T}_{s}$ and $\mathcal{T}_{v}$ as the basic distribution, denoted as $\mathcal{D}_{s}$ and $\mathcal{D}_{v}$, as shown in Fig. \ref{Fig_D_CIFAR10}- \ref{Fig_D_MIN}.

3). For the tracing results of $n, n\in[2,10]$ models, we conduct the sampling process (take one sample $S_{s}$ from $\mathcal{D}_{s}$ and $n-1$ sample $\mathbb{S}^{n-1}_v$ from $\mathcal{D}_{v}$) 10000 times. 

4) For each sampling, if $S_{s}>max(\mathbb{S}^{n-1}_v)$, we consider it as a correct tracing sample. We record the total number of correct tracing $N^{n}_C$ in 10000 samplings. The final tracing accuracy of $n$ models can be calculated with $N^{n}_C$/10000.

The results are shown in Fig. \ref{Fig_CIFAR10}-\ref{Fig_MIN}. The attack we choose is HSJA\cite{chen2020hopskipjumpattack}, and $\alpha$ is fixed as 0.15. It can be seen that with the increasing number of distributed copies, the tracing accuracy gradually decreases. But with 10 branches, it can still maintain more than 90\% accuracy for CIFAR10 and GTSRB. Besides, the estimated tracing performance is almost the same as the actual experiment results, which indicates the correctness of our analysis.

\section{Discussion}\label{Discussion}
\subsection{The importance of noise-sensitive loss}
In the proposed mechanism, making $\mathcal{T}_i$ easier to be attacked is the key for tracing. We design the noise-sensitive loss to meet the requirement. In this section, experiments will be conducted to show the importance of noise-sensitive loss. We use two randomly initialized tracers as the comparison to conduct the tracing experiment on 1000 adversarial images. The adversarial attack is set as HSJA\cite{chen2020hopskipjumpattack}, $\alpha$ is fixed as 0.15. The experimental results are shown in Table \ref{Randomtracer}.
\begin{table}[h]
\vspace{-5pt}
\centering

\scalebox{0.7}{
\begin{tabular}{c|cc|cc|cc}
\toprule[2pt]
\multirow{2}{*}{Attack} & \multicolumn{2}{c|}{CIFAR10}     & \multicolumn{2}{c|}{GTSRB}  & \multicolumn{2}{c}{mini-ImageNet} \\
        & ResNet18     & VGG16     & ResNet18     & VGG16  & ResNet50     & VGG19 \\
\midrule
Random      & $57.9\%$ & $62.4\%$ & $53.9\%$ & $57.0\%$ & $56.2\%$ & $59.8\%$\\
\midrule
Proposed      & $99.3\%$ & $97.9\%$ & $98.7\%$ & $98.3\%$ & $97.5\%$ & $95.4\%$\\
\bottomrule[2pt]
\end{tabular}}
\caption{The trace accuracy of HSJA attack with different $\mathcal{T}$.}
\label{Randomtracer}
\end{table}

It can be seen that without noise-sensitive loss, the tracing accuracy of the random initialized tracer only achieves 60\%, which is much lower than the proposed noise-sensitive tracer. This indicates that noise-sensitive loss is very important in realizing accurate tracing, only setting different parameters of tracer is not enough to trap the attack to result in specific features.
\subsection{Non-transferability and traceability}
The concept of traceability is related but not equivalent to non-transferability. 
A non-transferable adversarial example works only on the victim model it is generated from. Therefore, tracing such non-transferable example may be a straightforward task. On the other hand, a transferable sample may be generic enough to work on many copies/models. The task of tracing becomes more meaningful in this scenario. Our ability to trace a non-transferable example demonstrates that the process of adversarial attack introduces distinct traceable features which are unique to each victim model. In this sense, traceability can serve as a fail-safe property in defending adversarial attacks.
There are many defense methods can satisfy non-transferrability, but once the defense fails, the model will not be effectively protected. But our experimental results show that for the proposed method, even if the defense fails, we still have a certain probability to trace the attacked model, as shown in Table \ref{TraceAccuracyNT}. We use the data of ``ResNet-CIFAR10'' task with HSJA \cite{chen2020hopskipjumpattack} and QEBA \cite{li2020qeba} as examples to show the specific tracing results.

\begin{table}[h]
\centering

\vspace{-5pt}
\scalebox{0.8}{
\begin{tabular}{c|c|cccccc}
\toprule[2pt]
Attack & $\alpha$ & \textbf{NTr} & \textbf{NTr}(+) & \textbf{Tr} & \textbf{Tr}(+) & \textbf{Tr} Rate & \textbf{Total} Rate \\
\midrule
\multirow{3}{*}{HSJA}     & 0.05  & 672    &  672  &  328 &  313 &  95.43\%   &    98.50\%    \\
                          & 0.1   & 973    &  973   &  27 &  19 &    70.37\%    & 99.20\%       \\
                          & 0.15  & 993    &  993   &  7 &  0 &   0\%   &   99.30\%   \\ \midrule
\multirow{3}{*}{QEBA}     & 0.05  & 840    &  840   &  160 &  156 &    97.50\%   &   99.60\%   \\
                          & 0.1   & 879    &  879   &  121 &  118 &   97.52\%    &     99.70\%   \\
                          & 0.15  & 859    &  859   &  141 &  138 &   97.87\%    &     99.7\% \\ 
\bottomrule[2pt]
\end{tabular}
}
\caption{The trace accuracy of different attacks.}
\label{TraceAccuracyNT}
\end{table}

In Table \ref{TraceAccuracyNT}, \textbf{NTr} and \textbf{Tr} indicate the number of non-transferrable samples and transferrable samples respectively. \textbf{NTr}(+) and \textbf{Tr}(+) indicate the number of successful tracing samples. We can see that for QEBA with $\alpha=0.05$, $0.1$, and $0.15$, the traceability to transferrable samples is all keep at a high level which is greater than 97\%. As for HSJA, when $\alpha=0.05$, 328 samples can be transferred, and the traceability of transferrable examples achieves 95.43\%. When $\alpha=0.15$, although the traceability of transferrable examples decreases to 0\%, only 7 samples are transferrable. So the total tracing rate is still at a high level. In general, the proposed method either guarantees the high non-transferability or the high tracing accuracy for transferred samples.


\subsection{Limitations and adaptive attacks}
Although the proposed system maintains certain traceability in the buyers-seller setting, there are still some limitations that need to be addressed. For example, once the attacker finds a way to attack $\mathcal{C}$ and bypass $\mathcal{T}_i$, the tracing performance may degrade. But we found that attacking such system could be a challenging topic itself (in our setting) as the attackers do not have access to all other copies and thus are unable to avoid the differences that our tracer exploits. Besides, it seems a more adaptive attack also comes with ``cost''. For instance, the approach of attacking $\mathcal{C}$ and bypassing $\mathcal{T}_i$ would degrade the visual quality of the attack. So future work may be paid on how to evade the attack by utilizing such ``cost''.

\section{Conclusion}
This paper researches a new aspect of defending against adversarial attacks that is traceability of adversarial attacks. The techniques derived could aid forensic investigation of known attacks, and provide deterrence to future attacks in the buyers-seller setting. As for the mechanism, we design a framework which contains two related components (model separation and origin tracing) to realize traceability. For model separation, we propose a parallel network structure which pairs a unique tracer with the original classifier and a noise-sensitive training loss. Tracer model injects the unique features and ensures the differences between distributed models. As for origin tracing, we design an output-logits-based tracing mechanism. Based on this, the traceability of the attacked models can be realized when obtaining the adversarial examples. The experiment of multi-dataset and multi-network model shows that it is possible to achieve traceability through the adversarial examples.




\bibliography{aaai23}
\end{document}